# Measurement of the heat flux normalised spin Seebeck coefficient of thin films as a function of temperature

G. Venkat[1], C.D.W. Cox[2], A. Sola[3], V. Basso[3], K. Morrison[2]

[1]*Dept. of Materials Science and Engineering, University of Sheffield, Sheffield, United Kingdom, S10 2TG.*

[2]*Dept. of Physics, Loughborough University, Loughborough, United Kingdom, LE11 3TU.*

[3]*Instituto Nazionale di Ricerca Metrologica, Strada delle Cacce 91, 10135, Torino, Italy.*

The spin Seebeck effect (SSE) has generated interest in the thermoelectric and magnetic communities for potential high efficiency energy harvesting applications, and spintronic communities as a source of pure spin current. To understand the underlying mechanisms requires characterization of potential materials across a range of temperatures, however, for thin films the default measurement of an applied temperature gradient (across the sample) has been shown to be compromised by the presence of thermal resistances. Here, we demonstrate a method to perform low temperature SSE measurements where instead of monitoring the temperature gradient, the heat flux passing through the sample is measured using two calibrated heat flux sensors. This has the advantage of measuring the heat loss through the sample as well as providing a reliable method to normalise the SSE response of thin film samples. We demonstrate this method with an $SiO_2/Fe_3O_4/Pt$ sample, where a semiconducting-insulating transition occurs at the Verwey transition, $T_V$, of $Fe_3O_4$ and quantify the thermomagnetic response above and below $T_V$.

## I. INTRODUCTION

The spin Seebeck effect (SSE) was demonstrated in 2008 by Uchida *et al*[1] and has since been widely studied from the perspective of improving the efficiency of thermoelectric devices for energy harvesting.[2] While it was initially demonstrated in the transverse configuration in a metallic system such as Pt/NiFe[1], it was shortly thereafter measured in the longitudinal configuration in magnetic insulators such as Yttrium Iron Garnet (YIG), which removes parasitic effects like the planar Nernst effect in the SSE response,[3] as well as being ideally suited for energy harvesting applications. The SSE has subsequently been measured in a variety of systems such as ferromagnetic semiconductors[4], antiferromagnets[5], paramagnets[6] and ferrimagnets[7].

In the longitudinal configuration, the SSE involves application of a temperature gradient ($\nabla T$) normal to the plane of a thin film with the magnetization ($\boldsymbol{M}$) in the plane, which generates a spin current ($\boldsymbol{J}_{\text{s}}$) along the direction of $\nabla T$. For detection, $\boldsymbol{J}_s$ is then converted to a transverse charge current in a material with high spin-orbit coupling via the inverse spin Hall effect (ISHE). The voltage measured is then:

$$V_{\text{ISHE}} = \frac{1}{2}\left(V_{\text{sample}}\big|_{M_{\text{sat}}^{+}} - V_{\text{sample}}\big|_{M_{\text{sat}}^{-}}\right), \qquad (1)$$

where $M_{\text{sat}}^{+}$ and $M_{\text{sat}}^{-}$ are fields of positive and negative saturation and $V_{\text{sample}} = V_{\text{ISHE}} + V_{\text{S}}$ with $V_{\text{S}}$ being the background voltage due to the ordinary Seebeck effect. In most studies a heater is used to establish a temperature gradient ($\nabla T$) across the sample. The temperature difference ($\Delta T$), across the sample stack is usually measured with thermocouples in order to normalise $V_{\text{ISHE}}$ as[8,9] :

$$S_{\nabla T} = \frac{V_{\text{ISHE}}}{\nabla T L_y}, \qquad (2)$$

where $L_y$ is the contact separation, $\nabla T = \frac{\Delta T}{L_z}$, and $L_z$ is the sample thickness.

However, it has been shown that $S_{\nabla T}$ is an unreliable measure of the SSE in thin films due to the presence of thermal interface resistances across the sample stack[9]. To circumvent this problem, Sola *et al.* defined a heat flux based coefficient[10]:

$$S_{\mathrm{J_Q}} = \frac{V_{\mathrm{ISHE}}}{\mathrm{J_Q} L_y}, \qquad (3)$$

where $J_Q = \frac{Q}{A}$ is the heat flux passing through the sample ($Q$ is the heat applied and $A$ is the cross-sectional area of the sample). $S_{\mathrm{J_Q}}$ has been shown to be a more reliable measure of the SSE especially for thin films and has been used to quantify the SSE performance of YIG:Pt[8,9], $Fe_3O_4$:Pt[11] and $Co_2MnSi$:Pt[12]. This measurement has the advantage of being insensitive to thermal interface resistances in the measurement, as well as any thermal shunting across the thin film's substrate. Additionally, by multiplying equation (3) by the thermal conductivity of the magnetic film, the temperature normalised coefficient (defined in equation 2) can be obtained.

In reports of low temperature SSE measurements either $S_{\nabla T}$ is quoted or $S_{\mathrm{J_Q}}$ is estimated assuming that the power, $Q$, supplied to the heater passes through the sample (without loss)[13,14,15]. This is not reliable as low temperature thermal measurements are susceptible to multiple sources of heat loss and is compounded by the fact that it is often difficult to find low temperature heat flux sensors. Nevertheless, there is a need to measure $S_{\mathrm{J_Q}}$ down to low temperatures so that the interplay of different mechanisms (a metal-insulating layer may result in a shift from spin dependent- to magnon spin- Seebeck driven spin current)[2,11] or the impact of magnetic and structural phase transitions (on the signal) can be studied[5,6]. It is also important to quantify the performance of materials which show an increase in their SSE response at low temperatures[16].

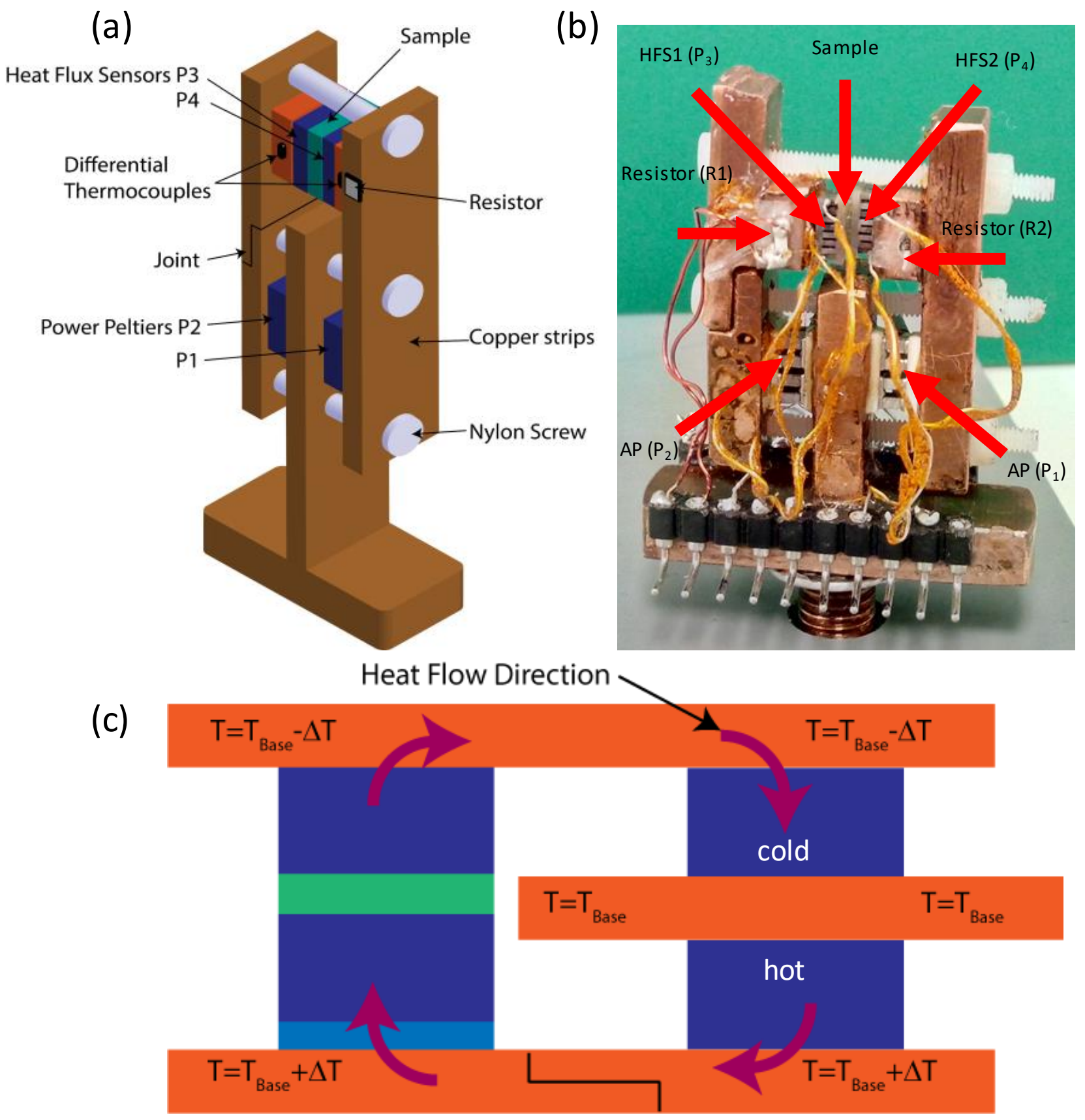

Figure 1 - The low temperature SSE (LTSSE) insert. (a) Schematic of the LTSSE holder showing the different components and (b) the corresponding photograph of the holder. (c) Schematic of the heat flow when setting up the temperature gradient using the activating Peltiers.

We describe here a low temperature SSE (LTSSE) measurement setup to measure $S_{\mathrm{JQ}}$ between 50 and 300 K using a closed cycle refrigerator (CCR) cryostat. The heat flux through the sample is measured with Peltier sensors on either side of the sample. Measuring $Fe_3O_4$ (80 nm):Pt (5 nm) thin films deposited on glass, we observe a decrease in $S_{\mathrm{JQ}}$ with temperature which we attribute to a general decrease in magnon accumulation.

## II Measurement setup

Fig. 1 (a) shows a schematic of the measurement setup. An oxygen free high thermal conductivity (OFHTC) copper mount was fabricated to attach to the cold head of an ARS CCR cryostat, where an indium seal was used between the cold head and the measurement puck to provide good thermal contact. The sample was held between two heat flux sensors (HFSs) which, in turn were attached to copper blocks. The design is based on a central pillar which is held at $T_{base}$, around which the thermal circuit is created (Fig. 1(c)). A thermal gradient can be established by two activating Peltier cells ($P_1$ and $P_2$) wired in series such that a hot side ($T_{base}+\Delta T$) is created on an adjoining copper strip and a cold side ($T_{base}- \Delta T$) on the second copper strip. The heat from the hot side then traverses the copper strip and through the copper block and sample stack as shown in Fig. 1(c). As two HFSs are used ($P_3$ and $P_4$), one on either side of the sample, a direct determination of the heat flux is obtained, and knowledge of the power generated by the heat source is not required. Whilst the activating Peltiers could provide enough heating power to establish a thermal circuit and detect an SSE signal at higher temperatures, two 0.5 kΩ resistor were also secured to the pillars to provide additional heating power if needed (and for calibration of the HFSs). The components and signal pins were secured to the copper holder using thermal epoxy (Stycast 1266) to ensure mechanical stability and good heat flow (Fig. 1(b)).

## <u>Calibration Procedure</u>

The calibration of the HFSs is important for precise determination of $J_Q$ and follows the procedure outlined by Sola *et al*[10]. The calibration was done in two modes: (a) heating from the top side(s) of the two HFSs, and (b) heating from between the two HFSs.

In mode (a) the resistors $R_1$ and $R_2$ (shown in Figure 1) were first used (in turn) to heat the HFSs from outside the stack such that an unknown heat, Q, passed through $P_3$ and $P_4$. This measurement enables determination of the relative sensitivity of each HFSs (to the other) and any difference between this measurement when heating with $R_1$ or $R_2$ is assumed to be due to heat loss through the HFSs. Upon the application of a heating current, the HFSs were allowed to equilibrate and settle (transient measurement), as shown in Figure 2(a). The value of the Peltier HFS voltage $V_p$ once equilibrium was reached (typically ~10-30 minutes) was determined for different heating currents supplied to $R_1/R_2$ and as a function of temperature. The ratio of the responses of $P_3$ and $P_4$, was then determined:

$$f = \frac{V_{P_4}}{V_{P_3}} \qquad (2)$$

and is shown in Figure 2(b). Note that the ~5% difference of the ratio measured for $R_1$ and $R_2$ is due to minor asymmetry in the heat losses from the HFSs. As we approach 300 K there is also some divergence, likely due to melting of the Apiezon N™ thermal grease used as the thermal interface between $P_3$ and $P_4$. With regards to asymmetry of the heat loss, heating from $R_1$ the expected voltage response could be described as:

$$V_{P3} = Q_{in}S_{P3} \qquad (3)$$

$$V_{P4} = (Q_{in}-Q_{loss})S_{P4} \qquad (4)$$

whereas heating from $R_2$:

$$V_{P4}' = Q_{in}S_{P4} \qquad (5)$$

$$V_{P3}' = (Q_{in}-Q_{loss})S_{P3} \qquad (6)$$

where $V_{P3}$ and $V_{P4}$ are the voltages measured from HFSs $P_3$ & $P_4$ respectively, $S_{P3}$ and $S_{P4}$ are the corresponding sensitivities (V/W), $Q_{in}$ is the heat that passes into the HFS assembly, and $Q_{loss}$ is the heat loss at each HFS. The difference in $V_{P4}/V_{P3}$ when heating with $R_1$ and $R_2$ can therefore be attributed to the asymmetry of $Q_{loss}$ with respect to $Q_{in}$, $V_{P3}$ and $V_{P4}$. Assuming that $Q_{loss}$ is small with respect to $Q_{in}$, the average of $V_{p4}/V_{p3}$ and $V_{p4}'/V_{p3}'$ will give the ratio of $S_{p4}$ to $S_{p3}$ as follows:

$$\frac{V_{P4}}{V_{P3}} = \frac{(Q_{in}-Q_{loss})}{Q_{in}}\frac{S_{p4}}{S_{p3}} \qquad (7)$$

$$\frac{V_{P4}'}{V_{P3}'} = \frac{Q_{in}}{(Q_{in}-Q_{loss})}\frac{S_{p4}}{S_{p3}} \qquad (8)$$

$$\left(\frac{V_{P4}}{V_{P3}} + \frac{V_{P4}'}{V_{P3}'}\right) = \frac{(Q_{in}-Q_{loss})}{Q_{in}}\frac{S_{p4}}{S_{p3}} + \frac{Q_{in}}{(Q_{in}-Q_{loss})}\frac{S_{p4}}{S_{p3}} = \frac{S_{p4}}{S_{p3}}\left(\frac{2Q_{in}{}^2-2Q_{in}Q_{loss}+Q_{loss}{}^2}{Q_{in}{}^2-Q_{in}Q_{loss}}\right). \qquad (9)$$

Where $Q_{loss}^2$ is considered negligible with respect to $2Q_{in}Q_{loss}$ and $2Q_{in}^2$, this reduces to:

$$\frac{1}{2}\left(\frac{V_{P4}}{V_{P3}} + \frac{V_{P4}'}{V_{P3}'}\right) = \frac{S_{p4}}{S_{p3}} \qquad (10)$$

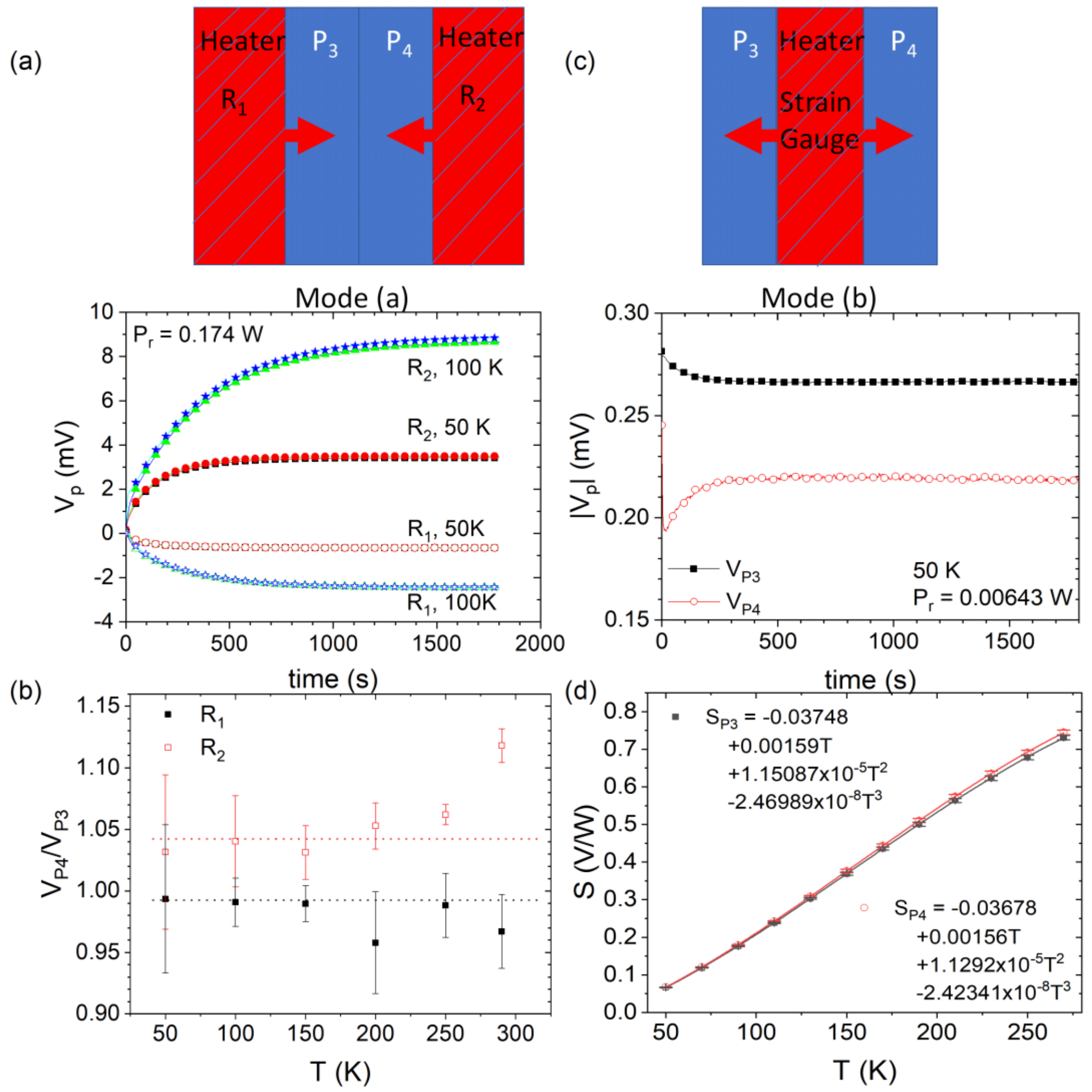

Figure 2 – Example data from Peltier calibrations. (a) The transient measurement signals from mode (a) where $P_3$ and $P_4$ are monitored whilst heated by $R_1$ (open symbols) or $R_2$ (closed symbols). (b) Corresponding ratio of $V_{P4}/V_{P3}$ as a function of temperature. (c) Example transient measurements from mode (b) where the total heat, Q, is known, but split between $P_3$ and $P_4$. (d) Summary of the calibration coefficients, $S_3$ and $S_4$ as a function of temperature (discussed in text). (The upper schematics show the heater configurations and directions of heat flow in each calibration mode).

Therefore, the average of the calibration measurements in mode (a) will give the ratio of HFS sensitivities, $S_{p4}/S_{p3}$. For example, for the data shown in Figure 2, $V_{p4}/V_{p3}$ and $V_{p4}'/V_{p3}'$ were 0.99 and 1.045, respectively. The average of this is 1.0175, which indicates that $S_{p4}$ = 1.0175 $S_{p3}$. Inserting into equations (7) or (8) indicate a $Q_{loss}$ of 2.6% (of $Q_{in}$). For comparison, for $Q_{in}$ = 1 W, this would give $Q_{loss}^2 = 6.835x10^{-4}$ $W^2$ compared to $2Q_{in}^2 = 2$ $W^2$ and $2Q_{in}Q_{loss} = 0.052$ $W^2$ (hence the assumption made for equation (10) holds). Mode (b) of the calibration requires heating of the HFSs by a resistor placed between them. In this scenario, Q is known, and is assumed to pass through either $P_3$ or $P_4$ with negligible loss. To limit heat loss from the heater wires (conductive) or sides (radiative), a thin film strain gauge was used. A series of heating currents were supplied, whilst the voltage drop across the strain gauge was monitored (4 wire measurement) so that the total power (Q) supplied to the 2 HFSs could be determined. An example of the transient responses from each HFS at 50 K is shown in Figure 2 (c). Notice that the response of $P_3$ is initially quite high, and decreases as equilibrium is reached, whilst the response of $P_4$ increases. This demonstrates the different timescales associated with each HFS due to the joint in the copper 'strip' seen in Figure 1(a). (This joint is a necessity to facilitate removal/insertion of the sample.) In other words, the difference in timescales is an indicator of the quality of the thermal path between $P_3$, $P_4$ and the cold finger. Given the placement of the heating resistor, if the heat flow through both HFSs is not equally split then the total power Q supplied by the strain gauge can be written as:

$$Q = \frac{|V_{P3}|}{S_{P3}} + \frac{|V_{P4}|}{S_{P4}} = \left(|V_{P3}| + {|V_{P4}|}/{f}\right)/S_{P3} \tag{11}$$

$$S_{P3} = \left(|V_{P3}| + {|V_{P4}|}/{f}\right)/Q \tag{12}$$

$$S_{P4} = (|V_{P3}|f + |V_{P4}|)/Q \tag{13}$$

where f is the ratio $S_{p4}/S_{p3}$ determined from calibration mode (a). Therefore, by applying equations (12) and (13) to the data obtained from calibration mode (b) alongside the ratio, f, determined from calibration mode (a), the sensitivities of HFSs $P_3$ and $P_4$ can be determined as a function of temperature, as shown in Figure 2(d).

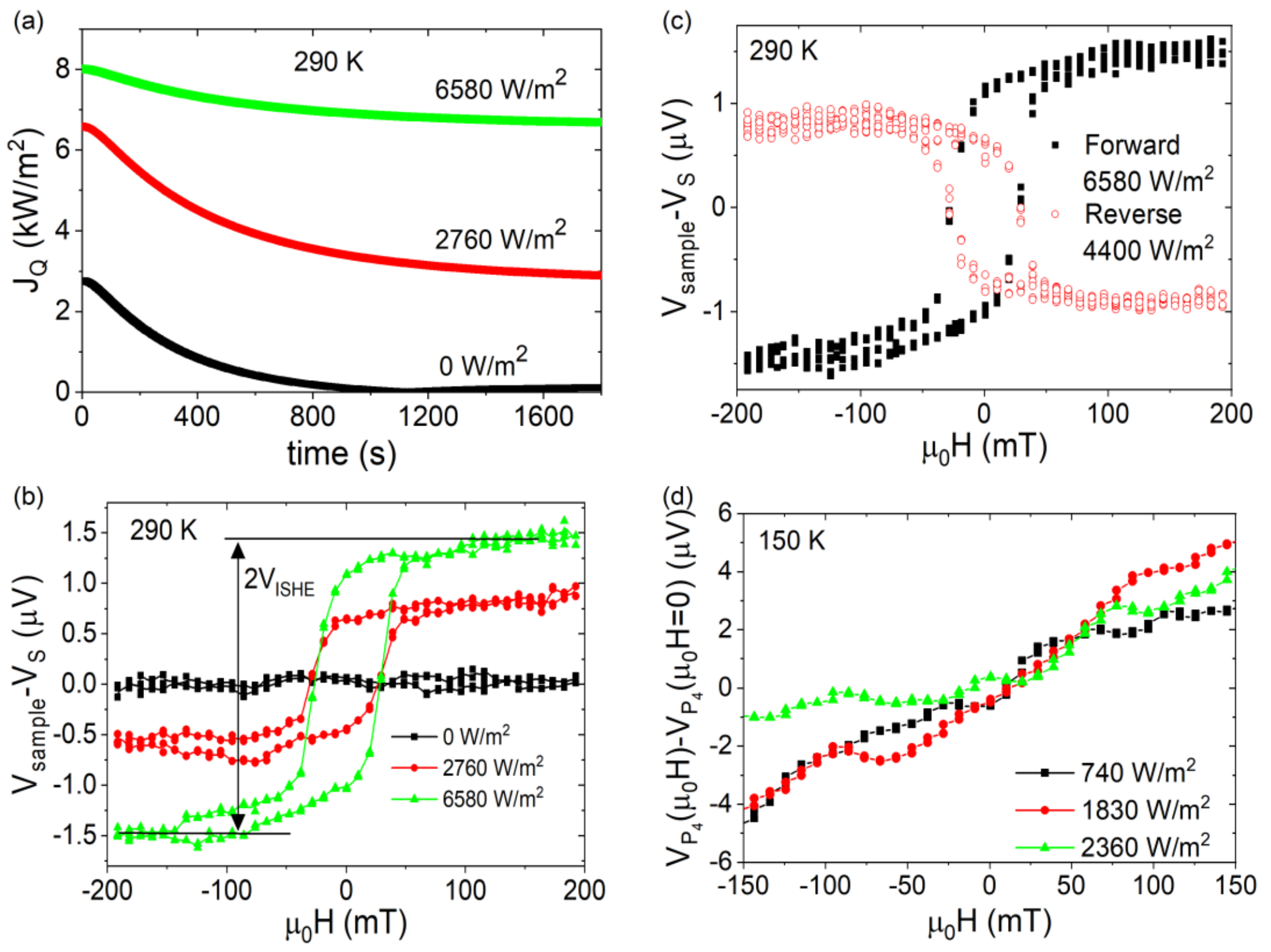

Figure 3 – Example SSE data as a function of applied magnetic field and heating option. (a) The transient heat flux through the sample relaxed by 30 minutes for each applied heat flux at 290 K. (Measurements were obtained with decreasing $J_Q$.) (b) $V_{sample}$-$V_s$ as a function of applied magnetic field for different measured heat fluxes at 290 K. The arrows show the calculation of $V_{ISHE}$ which is shown in Figure 4. (c) Observed reversal in sign of $V_{sample}$-$V_s$ when the temperature gradient direction is reversed at 290 K. (d) The variation of the HFS $P_4$ voltage with magnetic field. A 10 point average was taken to reduce the scatter in the data.

## Example Measurements

Once calibrated, several tests of the insert were run to assess the available heating power as a function of temperature (limiting to a base temperature of 50 K due to concerns of the stability of the HFSs to thermal cycling) and these are summarized in Figure 3. The sample initially considered was a 320 nm/5 nm $SiO_2/Fe_3O_4/Pt$ grown using PLD and described elsewhere[7,11]. The area of the sample and contact separation were $A = 26\ \text{mm}^2$ and $L_y = 3.5\ \text{mm}$, respectively. The heat flux across the sample was applied by heating and cooling the activating

Peltiers $P_1$ and $P_2$ using the thermal circuit described in Figure 1 (c). Figure 3 (a) shows the transient measurement as heat was applied to the sample (recorded by $P_3$). We can observe that the heat flux stabilized by 1800 s, which was the transient settle time used for all measurements. Figure 3 (b) shows the Seebeck subtracted sample voltage ($V_{sample}$-$V_S$) measured as a function of applied magnetic field ($\mu_0 H$) and different heat fluxes. The reversal in sign with field is a signature of the ISHE as the magnetization direction reverses and confirms that a magnetothermal signal is indeed being measured. Figure 3 (c) shows the reversal in sign of the data when heating from the other direction (i.e. reversal of the heat path shown in Figure 1 (c)). Note that due to the differing thermal paths between the HFSs and the cold finger, the available heating power (also indicated in Figure 1 (c)) is lower when heating from the P2 side. Once normalised according to equation (3) both datasets indicate $S_{\mathrm{J_Q}} = 47.5 \pm 2.4$ nVmW $^{-1}$. Finally, Figure 3 (d) shows the variation in the HFS signal as a function of applied magnetic field for different heat flux, where the value at $\mu_0 H$=0 has been subtracted from each dataset to make it easier to observe the field variation. This is shown for 150 K, as the increase in Johnson noise made it difficult to observe at higher temperatures. Note that for the $J_Q$ range studied here there is only a 0.02-0.04% variation in $V_{p4}$ when a field of 150 mT is applied.

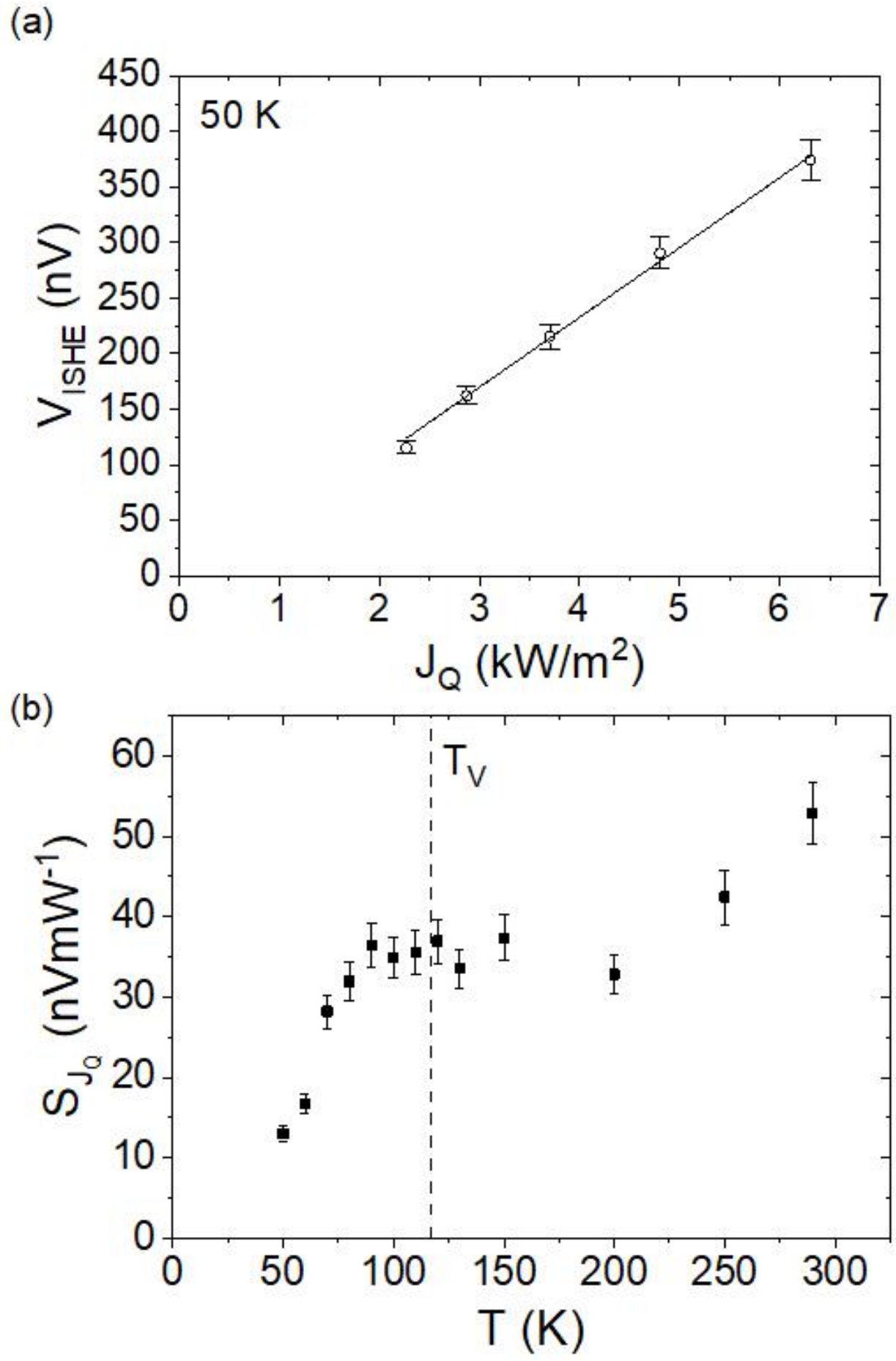

Figure 4 – Example SSE data for $Fe_3O_4$:Pt thin film. (a) $V_{ISHE}$ (shown by the arrows in Figure 3 (b)) as a function of heat flux, $J_Q$, at 50 K, where the gradient was used to determine the normalised coefficient, $S_{J_Q}$ defined in text. (b) Normalised coefficient as a function of temperature, where the position of the Verwey transition, $T_V$, is indicated.

Figure 4 shows a summary of example measurements of an 80 nm/5 nm $SiO_2/Fe_3O_4/Pt$ thin film as a function of temperature. The area of the sample and contact separation were $A = 27\ \mathrm{mm}^2$ and $L_y = 4.4$ mm, respectively. In (a), we show the variation of $V_{ISHE}$ with measured $J_Q$ and the slope of this is used to find the heat flux SSE coefficient $S_{\mathrm{J_Q}}$. Figure 4 (b) shows the variation of $S_{\mathrm{J_Q}}$ with temperature and the general decrease in $S_{\mathrm{J_Q}}$ with temperature is consistent with previous reports of $S_{\nabla T}$ for $Fe_3O_4$[17] and YIG[18,19,20]. This is attributed to the decrease in magnon population at lower temperatures. These thin films had a metal-insulator Verwey transition[21] temperature of 117 K[7] and we can see that whilst the SSE does not show an appreciable transition at this temperature, a plateau in the decrease of $S_{JQ}$ was observed. It is interesting to note that the peak observed between 50-100 K mirrors the variation of the thermal conductivity $\kappa$ previously seen for $Fe_3O_4$ thin films of comparable thickness[22] and this may lead to a flattening of the $S_{\nabla T}$ response from equation (2), which was observed by Ramos et.al[17].

In conclusion, we have developed an apparatus for simultaneously measuring the spin Seebeck response and heat flux passing through magnetic thin films from 300 K down to 50 K. The measurement involves accurately calibrating the performance of heat flux sensors and then measuring the inverse spin Hall signal from the sample as a function of applied heat flux and magnetic field at different sample temperatures in a cryostat. The simplicity of the measurement means that the spin Seebeck response can be accurately determined in thin films, whereby multiplying out by the thermal conductivity of the magnetic film, a coefficient normalised by the temperature gradient can be obtained. We hope that this measurement apparatus can contribute to progressing the state of the art in understanding the microscopic mechanism of the SSE in various material systems as well as developing spin Seebeck based energy harvesting devices.

**ACKNOWLEDGEMENTS**:

This work was supported by the EPSRC Fellow ship (EP/P006221/1). All supporting data will be made available via the Loughborough data repository via doi 10.17028/rd.lboro.11931669.